\documentclass[useAMS,onecolumn,usenatbib]{mn2e}

\usepackage{graphicx}
\usepackage{times}

\newcommand{\prt}{\partial}
\newcommand{\mrm}{\mathrm}

\title[Quasi-viscous accretion flow]
{Quasi-viscous accretion flow --- I: Equilibrium conditions and 
asymptotic behaviour}
\author[Bhattacharjee et al.]
{Jayanta K. Bhattacharjee,$^{1}$\thanks{jkb@bose.res.in}
Atri Bhattacharya,$^{2}$\thanks{atri@mri.ernet.in}
Tapas K. Das$^{2, 3}$\thanks{tapas@mri.ernet.in}
and Arnab K. Ray$^{4}$\thanks{akr@hbcse.tifr.res.in}\\
$^{1}$S. N. Bose National Centre for Basic Sciences, Sector III,
Block JD, Salt Lake, Kolkata 700098, India\\
$^{2}$Harish--Chandra Research Institute, Chhatnag Road, Jhunsi,
Allahabad 211019, India\\
$^{3}$Theoretical Institute for Advanced Research in Astrophysics,
101, Section 2, Kuang Fu Road, Hsinchu, Taiwan\\
$^{4}$Homi Bhabha Centre for Science Education, Tata Institute of
Fundamental Research, V. N. Purav Marg, Mankhurd, Mumbai 400088,
India}

\begin{document}


\maketitle

\label{firstpage}

\begin{abstract}
In a novel approach to studying viscous accretion flows, viscosity
has been introduced as a perturbative effect, involving a first-order
correction in the $\alpha$-viscosity parameter. This method reduces
the problem of solving a second-order nonlinear differential equation
(Navier-Stokes equation) to that of an effective first-order equation. 
Viscosity breaks down the invariance of the
equilibrium conditions for stationary inflow and outflow solutions,
and distinguishes accretion from wind. Under 
a dynamical systems classification, the only feasible critical 
points of this ``quasi-viscous" flow are saddle points and spirals. 
A linearised and radially propagating time-dependent perturbation 
gives rise to secular instability on large spatial 
scales of the disc. Further, on these same length scales, the 
velocity evolution equation of the quasi-viscous flow has been 
transformed to bear a formal closeness with Schr\"odinger's 
equation with a repulsive potential. Compatible with the transport 
of angular momentum to the outer regions of the disc, a 
viscosity-limited length scale has been defined for the full 
spatial extent over which the accretion process would be viable.
\end{abstract}


\section{Introduction}
\label{sec1}

The role of viscosity in the formation of accretion discs has, over
the years, been recognised to be of paramount 
importance~\citep{dlb69,ss73,pri81,fkr02}. The standard Keplerian
distribution of gaseous matter around a central accretor is determined
by viscosity, without which there could be no angular momentum transport
on outer length scales, and, therefore, no infall. Viscosity in a 
Keplerian disc also has a bearing on the time scale of the inward
radial drift of matter~\citep{fkr02}. So viscosity leaves its imprint
on accretion processes in more ways than one. While these facts are 
not a matter of doubt anymore, knowledge of the exact nature of 
viscosity still proves elusive. No help is also forthcoming from the
fact that the observables of an accretion disc have been theoretically
shown to be independent of viscosity~\citep{fkr02}. To explain the 
enhanced outward transport of angular momentum, and the accompanying
inflow rate, it has been variously suggested that turbulence, ordinarily
hydrodynamic, or even magnetohydrodynamic~\citep{bh98}, holds the key
to this as yet unsolved question. As a result much of the literature 
in accretion-related studies has been devoted to viscosity from one
perspective or the other~\citep{ss73,lt80,pri81,mkfo84,bmc86,acls88,
ny94,ct95,cal97,pa97,fkr02,ap03,cd04,bs05,unrs06,sand07,sharma08,
lanz08,sbk08} 

In contrast to a viscosity-driven accretion process, another 
model --- the sub-Keplerian low angular momentum inviscid flow --- has 
by now also become   
well-established in accretion studies~\citep{az81,fuk87,c89,nf89, 
skc90,ky94,yk95,par96,msc96,lyyy97,das02,dpm03,ray03a,bdw04,das04,
abd06,dbd06,crd06,gkrd07}. This is a particularly
expedient and simple physical system to investigate, and is 
considered especially appropriate in describing the 
the rotating flow in the innermost regions of the disc, very 
close to the event horizon of a black hole. Steady global solutions 
of inviscid 
axisymmetric accretion on to a black hole have been meticulously
studied over the years, and at present there exists an extensive
body of literature devoted to the subject, with especial emphasis
on the transonic nature of solutions, the multitransonic character
of the flow, formation of shocks, and the stability of global solutions 
under time-dependent linearised perturbations. 

Having stressed the usefulness of the inviscid model among researchers
in accretion astrophysics, it should also be recognised that this model 
has its own limitations. 
It is easy to understand that while the presence of angular momentum leads 
to the formation of an accretion disc in the first place, a physical 
mechanism should also be found for the outward transport of angular 
momentum, especially if its distribution is not sub-Keplerian (which,
for instance, is the case for strongly-coupled black hole binary 
systems). This should then make possible the inward drift of the accreting 
matter into the potential well of the accretor. It has already been 
mentioned that viscosity has been known 
all along to be a just such a physical means to effect infall, although
the exact presciption for viscosity in an accretion disc is still
a matter of much debate~\citep{pl95,fkr02}. What is well appreciated, 
however,
is that the viscous prescription should be compatible with an enhanced
outward transport of angular momentum. The very well-known $\alpha$
parametrisation of~\citet{ss73} is based on this principle. 

And so it transpires that on global scales --- especially on the
very largest scales of a non-sub-Keplerian disc --- the inviscid model 
will encounter
difficulties in the face of the fact that without an effective 
outward transport of angular momentum, the accretion process cannot 
be sustained globally. To address this issue, what is being 
introduced in this paper is the ``quasi-viscous" disc model. This model 
involves prescribing a very small first-order viscous correction in the 
$\alpha$-viscosity parameter of~\citet{ss73}, about the zeroth-order
inviscid solution. In doing this, a viscous generalisation of the 
inviscid flow can be logically extended to capture the important physical
properties of accretion discs on large length scales, without compromising 
on the fundamentally simple and elegant features of the inviscid model. 
This is the single most appealing aspect of the quasi-viscous disc model 
vis-a-vis many other standard models of axisymmetric flows which 
involve viscosity. To dwell on this point further, while Keplerian
discs explain infall processes satisfactorily, as far as the role
of viscosity is concerned, there is the difficulty that the net
force driving infall is practically zero (resulting from an exact
balance of the centrifugal effects against gravity). On the other
hand, while sub-Keplerian inviscid discs are free of this difficulty,
they do not account for any direct outward transport of angular
momentum\footnote{However, in the absence of any viscous transport 
of angular momentum, jets launched from accretion discs are supposed
to be the only outlet for the intrinsic angular momentum of the 
infalling matter~\citep{witjet}.} --- something that is also necessary 
to bring about infall.
The truth probably lies somewhere in between. The quasi-viscous model 
tries to address that possible area of convergence. While it accounts 
for an angular momentum transport, it also ensures that there is an 
effective force in the flow to drive the accretion process from an
outer boundary to the event horizon of a black hole accretor. 

From a most general fluid dynamical viewpoint, the effect of viscosity 
is described by a second-order nonlinear differential equation --- 
the Navier-Stokes equation~\citep{ll87}. The inviscid limit, on the 
other hand, is mathematically founded on Euler's equation, which is 
a first-order nonlinear differential equation. The quasi-viscous flow
here is based on a perturbative scheme about the inviscid conditions, 
and so the governing equation for this kind of viscous flow can be 
suitably approximated to a first-order equation. In fluid dynamics, 
this is not a particularly unusual mathematical expedient when it 
comes to accounting for viscosity~\citep{bdp93}. 

The immediate effect of viscosity on stationary flow solutions has
been to break down the invariance of the equilibrium conditions of
inflows and outflows, something that is otherwise preserved well in 
the inviscid limit. The equilibrium conditions of the quasi-viscous
flow have been precisely identified, and
the nature of the equilibrium points (critical points) has been 
discerned by devising a first-order autonomous dynamical system
from the flow equations. In this manner it has been shown that the 
possible critical points in the phase plot can be either saddle 
points, or spirals, or nodes. In the inviscid limit an earlier 
study has shown that only saddle points and centre-type points 
can exist~\citep{crd06}. Now centre-type
points are a limiting case of spirals~\citep{js99}. Since the 
quasi-viscous model represents a 
generalisation of the global inviscid flow, but at the same time
also implies that viscosity can be tuned to arbitrarily small 
values, a likely scenario that emerges is that the centre-type 
points (associated with inviscid flows) will become spirals on 
the inclusion of viscosity in the flow, however small. This should 
have various ramifications, especially about connecting multiple
transonic solutions through standing shocks~\citep{c89,das02}.  

An earlier work~\citep{br07} on the quasi-viscous flow, driven 
by the classical Newtonian potential, has revealed an instability
--- secular instability --- when the stationary flow solutions 
are subjected to small time-dependent perturbations. By analogy, 
exactly this kind of instability is also seen to develop in Maclaurin 
spheroids on the introduction of a kinematic viscosity to a first 
order~\citep{sc87}. Similar features in the quasi-viscous flow have 
also been demonstrated in this paper, but in this instance, under 
the pseudo-Schwarzschild generalisation. 

On large length scales the quasi-viscous flow displays some 
interesting asymptotic behaviour. Under highly subsonic conditions,
the pertinent flow equation (Navier Stokes equation) can be 
transformed mathematically into an equation that resembles 
Schr\"odinger's equation with a repulsive potential. This has been 
physically connected to the accumulation of angular momentum 
on large scales (a property of the disc that is very probably also
related to the growth of the secular instability on large length
scales), and a limiting scale of length has been derived from this 
condition. 

Finally, it should be worth stressing the fact that the entire
treatment presented here has been completely analytic, and to the
extent that this work purports to study pseudo-Schwarzschild flows, 
it has accounted for the use of any kind of generalised pseudo-Newtonian 
potential to drive the accretion process. This study is the first in
a series, in which quasi-viscous accretion around a rotating black
hole will also be taken up later. This will reveal the influence of 
black
hole spin angular momentum (Kerr parameter) on various equilibrium
and stability criteria for the flow.   

\section{The equations of the quasi-viscous axisymmetric flow and 
its equilibrium conditions}
\label{sec2}

For the thin disc, under the condition of hydrostatic equilibrium along 
the vertical direction~\citep{mkfo84,fkr02}, two of the relevant flow 
variables are the drift velocity, $v$, and the surface density, $\Sigma$. 
In the thin-disc approximation the latter has been defined by vertically 
integrating the volume density, $\rho$, over the disc 
thickness, $H(r)$. This gives $\Sigma \cong \rho H$, and in terms 
of $\Sigma$, the continuity equation is set down as 
\begin{equation}
\label{consig}
\frac{\prt \Sigma}{\prt t}+ \frac{1}{r} \frac{\prt}{\prt r}
\left(\Sigma vr \right) =0 .
\end{equation}

The axisymmetric accretion flow, driven by the gravitational field of 
a centrally located black hole, is described in terms of the Newtonian 
geometry of space and time with the help of what is known as a 
pseudo-Newtonian 
potential, $\phi (r)$. This paper will make use of such a general 
expression for the potential, and so the analytical results presented 
here will hold good under the choice of any pseudo-Newtonian potential. 
Assumption of the 
hydrostatic equilibrium in the vertical direction will give the 
condition 
\begin{equation}
\label{aitch0}
H = \frac{r}{\sqrt{\gamma}} \frac{c_{\mrm s}}{v_{\mrm K}}, 
\end{equation} 
in which the local speed
of sound, $c_{\mrm s}$, and the local Keplerian velocity, $v_{\mrm K}$, 
are, respectively, defined as 
$c_{\mrm s}^2 = \gamma P/\rho$ and 
$v_{\mrm K}^2 = r \phi^{\prime}$, with the pressure, $P$, itself 
being expressed in terms of a polytropic equation of state, 
$P= K \rho^{\gamma}$ (following this, the speed of sound may also be
given as $c_{\mrm s}^2 = \prt P/\prt \rho$). Written explicitly, the 
disc height can, therefore, be written as 
\begin{equation}
\label{aitch}
H = \left(\gamma K \right)^{1/2} \frac{\rho^{(\gamma -1)/2}
r^{1/2}}{\sqrt{\gamma \phi^{\prime}}} ,
\end{equation}
and with the use of this result, the continuity equation could then
be recast as 
\begin{equation}
\label{conrho}
\frac{\prt}{\prt t} \left[\rho^{(\gamma +1)/2}\right] 
+\frac{\sqrt{\phi^{\prime}}}{r^{3/2}} 
\frac{\prt}{\prt r} \left[\frac{\rho^{(\gamma +1)/2}vr^{3/2}}
{\sqrt{\phi^{\prime}}}\right]=0 .
\end{equation}

The general condition for the balance of specific angular momentum 
in the flow is given by~\citep{fkr02}, 
\begin{equation}
\label{angsig}
\frac{\prt}{\prt t}\left(\Sigma r^2 \Omega \right) + \frac{1}{r}
\frac{\prt}{\prt r} \left[ \left(\Sigma v r \right) r^2 \Omega \right]
= \frac{1}{2 \pi r} \left(\frac{\prt \mathcal{G}}{\prt r}\right) ,
\end{equation}
where $\Omega$ is the local angular velocity of the flow, while the 
torque is given as 
\begin{equation}
\label{torque}
{\mathcal G} = 2 \pi r \nu \Sigma r^2 
\left(\frac{\prt \Omega}{\prt r}\right) ,
\end{equation}
with $\nu$ being the kinematic viscosity associated with the flow. With 
the use of the continuity equation, as equation~(\ref{consig}) gives it,
and going by the~\citet{ss73} prescription for the kinematic viscosity,
$\nu = \alpha c_{\mrm s} H$, it would be easy to reduce 
equation~(\ref{angsig}) to the form~\citep{fkr02,ny94} 
\begin{equation}
\label{angrho}
\frac{1}{v}\frac{\prt}{\prt t}\left(r^2 \Omega \right)
+ \frac{\prt}{\prt r}\left(r^2 \Omega \right) = \frac{1}{\rho vrH}
\frac{\prt}{\prt r}\left[\frac{\alpha \rho H c_{\mrm s}^2 r^3}
{\Omega_{\mrm K}}\left(\frac{\prt \Omega}{\prt r}\right)\right] ,
\end{equation}
with $\Omega_{\mrm K}$ being defined from $v_{\mrm K}= r \Omega_{\mrm K}$. 

Going back to equation~(\ref{conrho}), a new variable is defined as 
$f= \rho^{(\gamma +1)/2}vr^{3/2}/{\sqrt{\phi^{\prime}}}$, whose 
steady value, as it is very easy 
to see from equation~(\ref{conrho}), can be closely identified with the 
constant matter flux rate. In terms of this new variable,
equation~(\ref{conrho}) can be modified as 
\begin{equation}
\label{coneff}
\frac{\prt}{\prt t} \left[\rho^{(\gamma +1)/2}\right]
+\frac{\sqrt{\phi^{\prime}}}{r^{3/2}}
\left(\frac{\prt f}{\prt r}\right) =0 ,
\end{equation}
while equation~(\ref{angrho}) can be rendered as 
\begin{equation}
\label{angeff}
\frac{1}{v}\frac{\prt}{\prt t}\left(r^2 \Omega \right)
+ \frac{\prt}{\prt r}\left(r^2 \Omega \right) =  
\frac{\alpha \gamma K}{f}
\frac{\prt}{\prt r}\left[f \left(\frac{f^2 \Omega_{\mrm K}}
{\rho^2 v^3} \right) \frac{\prt \Omega}{\prt r}\right] .
\end{equation}

The inviscid disc model is given by the requirement that 
$r^2 \Omega = \lambda$, in which $\lambda$ is the constant specific
angular momentum. The quasi-viscous disc that is being proposed 
here will introduce a first-order correction in terms involving 
$\alpha$, the~\citet{ss73} viscosity parameter, about the 
constant angular momentum solution. Mathematically this will be  
represented by the prescription of an effective specific angular
momentum,
\begin{equation}
\label{effecang}
\lambda_{\mrm{eff}}(r) = r^2 \Omega = \lambda + \alpha r^2 
\tilde{\Omega} ,
\end{equation}
with the form of $\tilde{\Omega}$ having to be determined from 
equation~(\ref{angeff}), under the stipulation that the dimensionless 
$\alpha$-viscosity parameter is much smaller than unity. This smallness 
of the quasi-viscous correction induces only very small changes on the 
constant angular momentum background, and, therefore, neglecting all 
orders of $\alpha$ higher than the first, and ignoring any explicit 
time-variation of the viscous correction term, the latter being a 
standard method adopted also for Keplerian 
flows~\citep{le74,ss76,pri81,fkr02}, 
the dependence of $\tilde{\Omega}$ on $v$ and $\rho$ is obtained as 
\begin{equation}
\label{tilomeg}
\tilde{\Omega} = - \frac{2 \lambda \gamma K}{r^2} 
\left[\frac{f^2 \Omega_{\mrm K}}{\rho^2 v^3 r^3} +
\int \frac{f^2 \Omega_{\mrm K}}{\rho^2 v^3 r^3} \left(\frac{1}{f}
\frac{\prt f}{\prt r}\right)\, {\mrm d}r \right] . 
\end{equation}
With $\tilde{\Omega}$ thus defined, it becomes possible under
stationary conditions to set down equation~(\ref{effecang}) in a
modified form as 
\begin{equation}
\label{statioang}
\lambda_{\mrm{eff}}(r) = \lambda - 2 \alpha \lambda 
\left(\frac{c_{\mrm s}^2}{vv_{\mrm K}} \right) .  
\end{equation}

Lastly, the equation for radial momentum balance in the flow will 
also have to be modified under the condition of quasi-viscous 
dissipation. This has to be done according to the scheme outlined
in equation~(\ref{effecang})
by which, the centrifugal term, $\lambda^2_{\mrm{eff}}(r)/r^3$, of 
the radial momentum balance equation, will have to be corrected 
upto a first order in $\alpha$. This will finally lead to the result
\begin{equation}
\label{radmom}
\frac{\prt v}{\prt t} + v \frac{\prt v}{\prt r}
+ \frac{1}{\rho}\frac{\prt P}{\prt r} + \phi^{\prime}(r)
- \frac{\lambda^2}{r^3} - 2 \alpha \frac{\lambda}{r^3}
\left(r^2 \tilde{\Omega} \right) =0 , 
\end{equation}
with $\tilde{\Omega}$ being given by equation~(\ref{tilomeg}), and $P$ 
being expressed as a function of $\rho$ with the help of a polytropic 
equation of state, as it has been mentioned earlier. The steady solution
of equation~(\ref{radmom}) is given as 
\begin{equation}
\label{statiorad}
v \frac{{\mrm d} v}{{\mrm d} r} + \frac{1}{\rho}
\frac{{\mrm d}P}{{\mrm d} r} + \phi^{\prime}(r) -\frac{\lambda^2}{r^3}
+ 4 \alpha \frac{\lambda^2}{r^3} 
\left(\frac{c_{\mrm s}^2}{vv_{\mrm K}} \right) = 0 ,
\end{equation}
whose first integral cannot be obtained analytically because of the
$\alpha$-dependent term. In the inviscid limit, though, the integral
is easily obtained. This case will be governed by conserved conditions, 
and its solutions have been well-known in accretion 
literature~\citep{c89,das02,dpm03}. They will either be open solutions
passing through saddle points or closed paths about centre-type points. 
The slightest presence of viscous dissipation, however, will radically
alter the nature of solutions seen in the inviscid limit, and it may be 
easily understood that solutions forming closed paths about centre-type
points, will, under conditions of small-viscous correction, be of the 
spiralling kind~\citep{lt80,mkfo84,ap03}. This state of affairs is 
appreciated very easily by the analogy of the simple harmonic oscillator.
In the undamped state the phase trajectories of the oscillator will,
very much like the solutions of the inviscid flow, be either closed 
paths about centre-type points or open paths through 
saddle points~\citep{js99}. 
With the presence of even very weak damping the closed paths change 
into spiralling solutions. A more detailed analysis in the regard 
will be carried out in Section~\ref{sec3}. 

\subsection{The fixed points for polytropic flows}
\label{subsec21}

The pressure, $P$, is prescribed by an equation of state for the
flow~\citep{sc39}.
As a general polytropic it is given as $P=K \rho^{\gamma}$, where
$K$ is a measure of the entropy in the flow and $\gamma$ is the
polytropic exponent. In terms of $\gamma$, the polytropic index, $n$,
is defined as $n = (\gamma -1)^{-1}$~\citep{sc39}. These definitions
are necessary to recast the first integral of equation~(\ref{conrho}),
which is easily obtained for stationary conditions. Using the relation
between $\rho$ and $c_{\mrm s}$, afforded by the polytropic condition, 
the final expression for the integral could be presented as
\begin{equation}
\label{conpol1st}
c_{\mathrm{s}}^{2(2n +1)} \frac{v^2 r^3}{\phi^{\prime}}
= \frac{\gamma}{4 \pi^2} \dot{\mathcal{M}}^2 , 
\end{equation}
where $\dot{\mathcal{M}} = (\gamma K)^n \dot{m}$~\citep{skc90}
with $\dot{m}$, an integration constant itself, being physically the
matter flow rate.

To obtain the critical points (the equilibrium points) of the flow, 
it should be necessary to combine both  
equations~(\ref{statiorad}) and (\ref{conpol1st}), along with the
polytropic definition of the equation of state,
to arrive ultimately at
\begin{equation}
\label{dvdrpol}
\left(v^2 - \beta^2 c_{\mrm{s}}^2 \right)
\frac{\mrm{d}}{\mrm{d}r}(v^2) = \frac{2 v^2}{r}
\left[ \frac{\lambda^2}{r^2} 
\left(1 - \frac{4 \alpha c_{\mrm{s}}^2}
{\sqrt{v^2 r \phi^{\prime}}} \right)
- r \phi^{\prime}
+ \frac{1}{2}\beta^2 c_{\mrm{s}}^2
\left(3 - r \frac{\phi^{\prime \prime}}{\phi^{\prime}}\right)\right] , 
\end{equation}
with $\beta^2 = 2(\gamma +1)^{-1}$. The critical points of the flow
will be given by the condition that the entire right hand side of
equation~(\ref{dvdrpol}) will vanish along with the coefficient of
${\mrm{d}}(v^2)/{\mrm{d}r}$. Explicitly written down,
following some rearrangement of terms, this will give the two
critical point conditions as,
\begin{equation}
\label{critconpol}
v_{\mrm{c}}^2 = \beta^2 c_{\mrm{sc}}^2
= 2\left[r_{\mrm{c}} \phi^{\prime}(r_{\mrm{c}})
- \frac{\lambda^2}{r_{\mrm{c}}^2} \right] \left[3 - r_{\mrm{c}}
\frac{\phi^{\prime \prime}(r_{\mrm{c}})}{\phi^{\prime}(r_{\mrm{c}})}
- \frac{8 \alpha \lambda^2 \beta^{-2}}{\sqrt{v_{\mrm c}^2 r_{\mrm c}^5
\phi^{\prime}(r_{\mrm c})}} \right]^{-1} , 
\end{equation}
with the subscript ``${\mrm{c}}$" labelling critical point values.

The roots of $r_{\mrm{c}}$ could be fixed in terms of $\gamma$, $\alpha$,
$\lambda$ and $\dot{\mathcal{M}}$ in the $r$ --- $v^2$ phase portrait
of the stationary flow. In order to do so, the latter condition in 
equation~(\ref{critconpol}) can be further modified with the help of 
the former condition to eliminate $v_{\mrm c}$. This will lead to
\begin{equation}
\label{critfixpol}
2\left[r_{\mrm{c}} \phi^{\prime}(r_{\mrm{c}})
- \frac{\lambda^2}{r_{\mrm{c}}^2} \right] \left[3 - r_{\mrm{c}}
\frac{\phi^{\prime \prime}(r_{\mrm{c}})}{\phi^{\prime}(r_{\mrm{c}})}
\right]^{-1} =
\beta^2 c_{\mrm{sc}}^2 -
\frac{8 \alpha \lambda^2 c_{\mrm{sc}}^2}{\sqrt{\beta^2 c_{\mrm{sc}}^2
r_{\mrm c}^5 \phi^{\prime}(r_{\mrm c})}} \left[3 - r_{\mrm{c}}
\frac{\phi^{\prime \prime}(r_{\mrm{c}})}{\phi^{\prime}(r_{\mrm{c}})}
\right]^{-1} , 
\end{equation}
which is a relation that gives $r_{\mrm c}$ as a function of $\gamma$,
$\alpha$, $\lambda$ and $c_{\mrm{sc}}$. To eliminate the dependence on
$c_{\mrm{sc}}$, it will be necessary to substitute $v$ 
in equation~(\ref{conpol1st}) by using the critical point conditions. 
This will give 
\begin{equation}
\label{fixcee}
c_{\mrm{sc}}^2 = \left[\frac{\gamma {\dot{\mathcal{M}}}^2 
\phi^{\prime}(r_{\mrm{c}})}{4 \pi \beta^2 r_{\mrm c}^3}\right]^{1/2(n+1)} , 
\end{equation}
a result with the obvious implication that the dependence of 
$r_{\mrm{c}}$ will finally be given as 
$r_{\mrm{c}}= f_1(\gamma, \alpha, \lambda, \dot{\mathcal{M}})$. One 
very interesting consequence of the presence of viscosity in 
equation~(\ref{critfixpol}) is that the sign of the square root on the 
right hand side has to be chosen according to whether one is studying
inflow solutions or outflow solutions. For inflows a negative sign 
would have to be extracted from the square root, while for outflows 
the chosen sign would have to be positive. This suggests
that the invariance of the coordinates of the critical points in the 
$r$ --- $v^2$ plane would be lost because of dissipation, as opposed
to the fully conserved inviscid case~\citep{crd06}. Viscosity, therefore,
will distinguish accretion solutions from wind solutions. 
 
The slope of the continuous solutions which could possibly pass through
the critical points are to be obtained by applying the L'Hospital rule
on equation~(\ref{dvdrpol}) at the critical points. This will give a 
quadratic equation for the slope of stationary solutions at the critical 
points themselves in the $r$ --- $v^2$ phase portrait. The resulting 
expression will read as  
\begin{equation}
\label{lhosp}
\left[\frac{\mrm{d}}{\mrm{d}r}(v^2){\bigg{\vert}}_{\mrm c}\right]^2
+ \left({\mathcal{Z}}_1 + \alpha {\mathcal{Z}}_2 \right)
\left[\frac{\mrm{d}}{\mrm{d}r}(v^2){\bigg{\vert}}_{\mrm c}\right]
+ \left({\mathcal{Z}}_3 + \alpha {\mathcal{Z}}_4 \right) = 0 , 
\end{equation} 
in which the constant coefficients, $\mathcal{Z}_1$, $\mathcal{Z}_2$,
$\mathcal{Z}_3$ and $\mathcal{Z}_4$ are given by 
\begin{displaymath}
\label{zed1}
{\mathcal Z}_1 =\frac{2}{\gamma}\left(\frac{\gamma -1}{\gamma +1}\right) 
\frac{c_{\mrm{sc}}^2}{r_{\mrm c}} 
\left[3 - r_{\mrm{c}}
\frac{\phi^{\prime \prime}(r_{\mrm{c}})}{\phi^{\prime}(r_{\mrm{c}})}
\right] ,
\end{displaymath}
\begin{displaymath}
\label{zed2}
{\mathcal Z}_2 =
- \left(\frac{3 - \gamma}{\gamma}\right)
\frac{2 \lambda^2 c_{\mrm{sc}}^2}{\sqrt{v_{\mrm{c}}^2
r_{\mrm c}^7 \phi^{\prime}(r_{\mrm c})}} , 
\end{displaymath}
\begin{displaymath}
\label{zed3}
{\mathcal Z}_3 = \frac{c_{\mrm{sc}}^2}{\gamma}\left[\frac{6 \lambda^2}
{r_{\mrm c}^4} + 2 \phi^{\prime \prime}(r_{\mrm{c}}) + \frac{2}{\gamma +1}
c_{\mrm{sc}}^2 \left\{ \frac{3}{r_{\mrm c}^2} + 
\frac{\mrm{d}}{\mrm{d}r}\left(\frac{\phi^{\prime \prime}}{\phi^{\prime}}
\right) {\bigg{\vert}}_{\mrm c} \right\} \right]
\end{displaymath}
and
\begin{displaymath}
\label{zed4}
{\mathcal Z}_4 = 
 \frac{8 \lambda^2 c_{\mrm{sc}}^2}{\sqrt{v_{\mrm{c}}^2
r_{\mrm c}^7 \phi^{\prime}(r_{\mrm c})}} 
\frac{c_{\mrm{sc}}^2}{\gamma}
\left[\sqrt{r_{\mrm c}^7 
\phi^{\prime}(r_{\mrm c})} \frac{\mrm{d}}{\mrm{d}r} \left\{
\left(r^7 \phi^{\prime}\right)^{-1/2} \right\} {\bigg{\vert}}_{\mrm c}
+ \left( \frac{\gamma -1}{\gamma +1} \right) \frac{1}{r_{\mrm c}}
\left\{3 - r_{\mrm{c}}
\frac{\phi^{\prime \prime}(r_{\mrm{c}})}{\phi^{\prime}(r_{\mrm{c}})}
\right\} \right] , 
\end{displaymath}
respectively. 

\subsection{The fixed points for isothermal flows}
\label{subsec22}

For an isothermal flow, the appropriate equation of state is given 
by $P= \rho {\kappa}T/\mu m_{\mrm{H}}$, in which $\kappa$ is 
Boltzmann's constant, $T$ is the constant temperature, 
$m_{\mrm{H}}$ is the mass of a hydrogen atom and $\mu$ is the 
reduced mass, respectively. The definition for $H$ in 
equation~(\ref{aitch0}) will have to be modified slightly by 
setting $\gamma =1$~\citep{ap03,crd06}. 
The local speed of sound will also be modified to become a 
global constant of the flow, going by the definition  
$c_{\mrm s}^2 = \prt P/\prt \rho$. 

Going back to equation~(\ref{statiorad}) and using the linear dependence 
between $P$ and $\rho$ as the appropriate equation of state, will lead to 
\begin{equation}
\label{statioradiso}
v \frac{{\mrm d} v}{{\mrm d} r} + c_{\mrm s}^2  
\frac{{\mrm d}}{{\mrm d} r}\left(\ln \rho \right) + \phi^{\prime}(r) 
-\frac{\lambda^2}{r^3}
+ 4 \alpha \frac{\lambda^2}{r^3}
\left(\frac{c_{\mrm s}^2}{vv_{\mrm K}} \right) = 0 ,
\end{equation}
while the first integral of the stationary continuity condition, 
given by equation~(\ref{conrho}) will give   
\begin{equation}
\label{coniso1st}
\frac{\rho^2 v^2 r^3}{\phi^{\prime}}
= \frac{\dot{m}^2}{4 \pi^2 c_{\mrm{s}}^2} . 
\end{equation}
As it has been done for polytropic flows, the two foregoing equations 
can be combined to obtain
\begin{equation}
\label{dvdriso}
\left(v^2 -  c_{\mrm{s}}^2 \right)
\frac{\mrm{d}}{\mrm{d}r}(v^2) = \frac{2 v^2}{r}
\left[ \frac{\lambda^2}{r^2} 
\left(1 - \frac{4 \alpha c_{\mrm{s}}^2}
{\sqrt{v^2 r \phi^{\prime}}} \right)
- r \phi^{\prime}
+ \frac{1}{2} c_{\mrm{s}}^2
\left(3 - r \frac{\phi^{\prime \prime}}{\phi^{\prime}}\right)\right] , 
\end{equation}
from which the critical point conditions are easily identified as
\begin{equation}
\label{critconiso}
v_{\mrm{c}}^2 = c_{\mrm{s}}^2
= 2\left[r_{\mrm{c}} \phi^{\prime}(r_{\mrm{c}})
- \frac{\lambda^2}{r_{\mrm{c}}^2} \right] \left[3 - r_{\mrm{c}}
\frac{\phi^{\prime \prime}(r_{\mrm{c}})}{\phi^{\prime}(r_{\mrm{c}})}
- \frac{8 \alpha \lambda^2}{\sqrt{v_{\mrm c}^2 r_{\mrm c}^5
\phi^{\prime}(r_{\mrm c})}} \right]^{-1} . 
\end{equation}
Fixing the critical point is much simpler a task for isothermal flows. 
As it has been done for the polytropic case, $v_{\mrm c}$ has to be 
eliminated first from the second critical point condition 
in equation~(\ref{critconiso}), to obtain 
\begin{equation}
\label{critfixiso}
2\left[r_{\mrm{c}} \phi^{\prime}(r_{\mrm{c}})
- \frac{\lambda^2}{r_{\mrm{c}}^2} \right] \left[3 - r_{\mrm{c}}
\frac{\phi^{\prime \prime}(r_{\mrm{c}})}{\phi^{\prime}(r_{\mrm{c}})}
\right]^{-1} =
c_{\mrm s}^2 -
\frac{8 \alpha \lambda^2 c_{\mrm s}^2}{\sqrt{c_{\mrm s}^2
r_{\mrm c}^5 \phi^{\prime}(r_{\mrm c})}} \left[3 - r_{\mrm{c}}
\frac{\phi^{\prime \prime}(r_{\mrm{c}})}{\phi^{\prime}(r_{\mrm{c}})}
\right]^{-1} .
\end{equation}
In this expression, the speed of sound, $c_{\mrm{s}}$, is
globally constant, and so having arrived at the critical point
conditions, it should be easy to see that $r_{\mrm{c}}$ and
$v_{\mrm{c}}$ have already been fixed in terms of a global
constant of the system. The speed of sound can further be written
in terms of the temperature of the system as
$c_{\mrm{s}} = \Theta T^{1/2}$, where
$\Theta = (\kappa/\mu m_{\mrm H})^{1/2}$, and, therefore,
it should be entirely possible to give a functional dependence for
$r_{\mrm{c}}$, as $r_{\mrm{c}} = f_2(\alpha, \lambda, T)$. 
The slope of the solutions passing through the critical points in 
the $r$ --- $v^2$ phase portrait is obtained simply by setting 
$\gamma =1$ in equation~(\ref{lhosp}). 

\section{The character of the fixed points: A dynamical systems study} 
\label{sec3}

The stationary fluid equations describing a viscous flow are in
general second-order nonlinear differential equations~\citep{ll87}. 
There is as yet no standard prescription for deriving analytic solutions
from these equations. Therefore, for any understanding of the behaviour
of the flow solutions, a numerical integration is in most cases
the only recourse. On the other hand, an alternative approach
could be made to this question, if the governing equations are
set up to form a standard first-order dynamical system~\citep{js99}.
The mathematical formalism of the stationary quasi-viscous flow is 
premised on the two first-order differential equations, given by 
equations~(\ref{statiorad}) and (\ref{conpol1st}). Of these, the 
former equation is the result of an approximation to obtain an 
appropriate first-order differential equation to describe a viscous
flow. This kind of an approximation is quite usual in general 
fluid dynamical studies~\citep{bdp93},
and short of carrying out any numerical integration, this approach
allows for gaining physical insight into the behaviour of the
flows to a surprising extent. As a first step towards this end, for 
the stationary polytropic flow, as given by equation~(\ref{dvdrpol}),
it should be necessary to parametrise this equation and set up
a coupled autonomous first-order dynamical system as~\citep{js99}
\begin{eqnarray}
\label{dynsys}
\frac{\mrm{d}}{\mrm{d}\tau}(v^2)&=& 2v^2 \left[
\frac{\lambda^2}{r^2} 
\left(1 - \frac{4 \alpha c_{\mrm{s}}^2}
{\sqrt{v^2 r \phi^{\prime}}} \right)
- r \phi^{\prime} + \frac{1}{2}
\beta^2 c_{\mrm{s}}^2 \left( 3 - r \frac{\phi^{\prime \prime}}
{\phi^{\prime}} \right) \right] \nonumber \\
\frac{\mrm{d}r}{\mrm{d} \tau}&=& r \left(v^2 -
\beta^2 c_{\mrm{s}}^2 \right) , 
\end{eqnarray}
in which $\tau$ is an arbitrary mathematical parameter. With respect
to accretion studies in particular, this kind of parametrisation
has been reported before~\citep{bmc86,rb02,ap03,crd06,mrd07,gkrd07}.
This opens the way to explore the mathematical nature of the critical 
points much more thoroughly. 

The critical points (which give the equilibrium conditions in the
flow) have themselves been fixed in terms of the flow
constants. About these fixed point values, upon using a perturbation
prescription of the kind
$v^2 = v_{\mrm{c}}^2 + \delta v^2$, $c_{\mrm{s}}^2 =
c_{\mrm{sc}}^2 + \delta c_{\mrm{s}}^2$ and
$r = r_{\mrm{c}} + \delta r$, it becomes possible to derive a set 
of two autonomous first-order linear differential equations in the
$\delta r$ --- $\delta v^2$ plane, with $\delta c_{\mrm{s}}^2$ itself
being expressed in terms of $\delta r$ and $\delta v^2$
from equation~(\ref{conpol1st}) as
\begin{equation}
\label{varsound}
\frac{\delta c_{\mrm{s}}^2}{c_{\mrm{sc}}^2} = - \frac{\gamma -1}
{\gamma + 1} \left[ \frac{\delta v^2}{v_{\mrm{c}}^2}
+ \left\{ 3 - r_{\mrm{c}}\frac{\phi^{\prime \prime}(r_{\mrm{c}})}
{\phi^{\prime}(r_{\mrm{c}})} \right\}
\frac{\delta r}{r_{\mrm{c}}}
\right] . 
\end{equation}
The resulting coupled set of linear equations in $\delta r$ and
$\delta v^2$ will be given as
\begin{eqnarray}
\label{lindynsys}
\frac{\mrm{d}}{\mrm{d}\tau}(\delta v^2) &=& A \delta v^2
+ B \delta r \nonumber \\
\frac{\mrm{d}}{\mrm{d}\tau}(\delta r) &=& C \delta v^2 
+ D \delta r , 
\end{eqnarray}
in which the constant coefficients $A$, $B$, $C$
and $D$ are to be read as 
\begin{displaymath}
\label{coae}
A = \left(\frac{\gamma -1}{\gamma +1}\right)
{\mathcal X}v_{\mrm c}^2 + \left(\frac{3 \gamma -1}{\gamma +1}\right)
\frac{4 \alpha \lambda^2 c_{\mrm{sc}}^2}{\sqrt{v_{\mrm c}^2
r_{\mrm c}^5 \phi^{\prime}(r_{\mrm c})}} , 
\end{displaymath}
\begin{displaymath}
\label{cobee}
B = -2 v_{\mrm c}^2 
\left [\frac{2 \lambda^2}{r_{\mrm{c}}^3} +
\phi^{\prime}(r_{\mrm{c}}) + r_{\mrm{c}}\phi^{\prime \prime}
(r_{\mrm{c}}) + \frac{\beta^2}{2}
\frac{\phi^{\prime \prime}(r_{\mrm{c}})}{\phi^{\prime}(r_{\mrm{c}})}
{c_{\mrm{sc}}^2} \mathcal{Y} + \frac{\beta^2}{2}
\left(\frac{\gamma -1}{\gamma + 1} \right)
\frac{{c_{\mrm{sc}}^2}}{r_{\mrm{c}}}
{\mathcal X}^2 \right]
\end{displaymath}
\begin{displaymath}
\label{cobee1}
\qquad \qquad 
+ \frac{8\alpha \lambda^2 v_{\mrm c}^2 c_{\mrm{sc}}^2}{\sqrt{v_{\mrm c}^2
r_{\mrm c}^7 \phi^{\prime}(r_{\mrm c})}} 
\left[\left(\frac{\gamma -1}{\gamma +1}\right) 
 \left\{ 3 - r_{\mrm{c}}\frac{\phi^{\prime \prime}(r_{\mrm{c}})}
{\phi^{\prime}(r_{\mrm{c}})} \right\} + \frac{5}{2} +
\frac{r_{\mrm c} \phi^{\prime \prime}(r_{\mrm{c}})}
{2 \phi^{\prime}(r_{\mrm{c}})} \right] , 
\end{displaymath}
\begin{displaymath}
\label{cocee}
C = \left(\frac{2 \gamma}{\gamma +1}\right) r_{\mrm c}
\end{displaymath}
and 
\begin{displaymath}
\label{codee}
D = - \left(\frac{\gamma -1}{\gamma +1}\right)
{\mathcal X}v_{\mrm c}^2 ,
\end{displaymath}
under the further definition that
\begin{displaymath}
\label{coeffx}
\mathcal{X} = r_{\mrm{c}}\frac{\phi^{\prime \prime}(r_{\mrm{c}})}
{\phi^{\prime}(r_{\mrm{c}})} - 3 ,  
\end{displaymath}
and 
\begin{displaymath}
\label{coeffy}
\mathcal{Y} = 1 + r_{\mrm{c}}
\frac{\phi^{\prime \prime \prime}(r_{\mrm{c}})}
{\phi^{\prime \prime}(r_{\mrm{c}})}
- r_{\mrm{c}}\frac{\phi^{\prime \prime}(r_{\mrm{c}})}
{\phi^{\prime}(r_{\mrm{c}})} . 
\end{displaymath}

Trying solutions of the type $\delta v^2 \sim \exp(\Omega \tau)$
and $\delta r \sim \exp(\Omega \tau)$ in equations~(\ref{lindynsys}), 
will deliver the eigenvalues $\Omega$, which are the growth rates of 
$\delta v^2$ and $\delta r$, as 
\begin{equation}
\label{eigen}
\Omega^2 - \left(A + D \right) \Omega 
+ \left(AD - BC\right) = 0 . 
\end{equation}
Under a further definition that $\mathrm{P} = A + D$, 
$\mathrm{Q} = AD - BC$ and 
$\Delta = {\mathrm{P}}^2 - 4 \mathrm{Q}$, 
the solution of the foregoing quadratic equation can be written as 
\begin{equation}
\label{eigenquad}
\Omega = \frac{{\mathrm P} \pm \sqrt{\Delta}}{2} . 
\end{equation}

Once the 
position of a critical point, $r_{\mrm{c}}$, has been ascertained, it 
is then a straightforward task to find the nature of that critical point 
by using $r_{\mrm{c}}$ in equation~(\ref{eigenquad}) and all its 
associated values.
Since it has been discussed in Section~\ref{sec2} that $r_{\mrm{c}}$
is a function of $\alpha$, $\lambda$ and $T$ for isothermal flows, and 
a function of $\gamma$, $\alpha$, $\lambda$ and $\dot{\mathcal M}$ for 
polytropic flows, it effectively implies that $\Omega$ can, in principle, 
be rendered as a function of the flow parameters for either kind of flow.
For isothermal flows, starting from equation~(\ref{dvdriso}),
a similar expression for the related eigenvalues may likewise be derived.
The algebra in this case is much simpler and it is easy to show that
for isothermal flows the relevant results could be derived by simply
setting $\gamma = 1$ in equations~(\ref{lindynsys}).

The nature of the possible critical points can also be predicted from 
the form of $\Omega$ in equation~(\ref{eigenquad}). If $\Delta > 0$, 
then a critical point can be either a saddle or a node~\citep{js99}. 
The precise nature of the critical point will then be dependent on 
the sign of $\mathrm Q$. If $\mathrm{Q} < 0$, then the critical point
will be a saddle point. Such points are always notoriously unstable 
in terms of the sensitivity in generating a solution through them, 
after starting from a boundary value far away from the critical 
point~\citep{rb02,rbcqg07,rnr07}. On the other hand, if $\mathrm{Q} >0$, 
then the critical point will be a node. Such a point may or may not be 
stable, depending on the sign of $\mathrm{P}$. If $\mathrm{P} <0$,
then the node will be stable. 

A completely different class of critical points will result when 
$\Delta <0$. These points will be like a spiral (a focus). Once 
again, the stability of the spiral will depend on the sign of 
$\mathrm P$. If $\mathrm P <0$, then the spiral will be stable. 
For inflow solutions in the quasi-viscous disc, the form of $\mathrm P$
(deriving from the sum of $A$ and $D$) shows that it will {\em always}
be negative. This is because for inflows the negative root of 
$v_{\mrm c}$ has to be extracted (i.e. $v_{\mrm c} <0$) from the 
square root in the definition of $A$. Which will obviously mean that
if the critical point is either a spiral or a node, then it will be
stable, with flow solutions in the neighbourhood of the critical 
point converging towards it. 

The quasi-viscous prescription is based on the requirement that 
viscosity will only have a small perturbative effect about the 
conserved inviscid flow. In other words, one could tune the 
viscosity parameter, $\alpha$, to arbitrarily small (but
non-zero) values. In this kind of a situation it is much more likely
than not that $\Delta <0$, and that the stable critical point will
be a spiral (nodal points, however, cannot be ruled out completely,
as~\citet{ap03} have shown). Therefore, the most likely picture that
emerges as far as the phase portrait of the flow is concerned, is 
that there will be adjacent unstable saddle points and stable
spiral points (adjacent points cannot be both stable or unstable
simultaneously). This argument is in keeping with an earlier 
study~\citep{crd06} on the inviscid disc, 
where a generic conclusion that was drawn about the critical points 
was that for a conserved pseudo-Schwarzschild axisymmetric flow 
driven by any potential, the only admissible critical points would 
be either saddle points or centre-type points. For a saddle point, 
$\Omega^2 > 0$, while for a centre-type point, $\Omega^2 < 0$, with 
$\Omega^2$ being real on both occasions.
Noting that a centre-type point is merely a special case 
($\mathrm P =0$) of a spiral, introduction of viscosity as a 
small perturbative effect certainly represents a physical 
generalisation. But with this, what is lost from the phase portrait
of the flow are homoclinic trajectories connecting a saddle point
to itself, or even heteroclinic trajectories connecting two saddle
points, although one might still argue that heteroclinic paths will 
exist to connect saddle points with spirals. 

Once the behaviour of all the physically relevant
critical points has been understood in this way, a complete qualitative
picture of the flow solutions passing through these points (if they
are saddle points), or in the neighbourhood of these points (if they
are spiral points), can be constructed, along with an impression
of the direction that these solutions can have in the phase portrait
of the flow~\citep{js99}. So what does that imply for multitransonicity,
especially about flow solutions which can be generated very far away
from the black hole accretor to reach its event horizon eventually?
Many earlier studies~\citep{c89,skc90,das02,dpm03,crd06} have taken
up this question in great detail, and it has been shown that for 
certain parameter-space values pertaining to the inviscid disc, 
three critical points can result. These are located in such a manner
that a centre-type point is flanked by two saddle points~\citep{skc90}
through which transonic solutions pass. 
For very small values of viscosity, it is now conceivable that the
centre-type point in the middle will become a stable spiral. This is
in fact very much in keeping with the conclusion of~\citet{lt80} that
the number of independent transonic solutions cannot exceed one plus
the number of spiral singularities. \citet{lt80} have further suggested
that in realistic physical situations, models with spiral singularities
are unstable and the critical transonic solution whenever it exists 
is unique in relevant situations. While the instability of the 
quasi-viscous disc will indeed be verified in Section~\ref{sec4}, 
the argument of~\citet{lt80} also conceivably has a strong bearing 
on another feature that is very intimately connected to multitransonicity
in accretion flow --- shocks, with or without 
dissipation~\citep{c89,skc90,das02,dpm03,cd04,sand07,fk07,lanz08}. 

\section{Travelling-wave perturbative analysis: Secular instability}
\label{sec4}

Many earlier works have taken up the question of the
stability of viscous thin disc
accretion~\citep{le74,ss76,ls77,kato78,us05}.
Regarding quasi-viscous accretion discs in particular, 
an earlier study~\citep{br07} has shown that stationary flow solutions
driven by the simple Newtonian potential suffer from an instability
under time-dependent perturbations --- both as a standing wave and as 
a high-frequency travelling wave. It need not always be
true that standing and travelling waves will simultaneously exhibit the
same qualitative properties as far as stability is concerned. Many
instances in fluid dynamics bear this out. In the case of binary
fluids, standing waves indicate instability, as opposed to travelling
waves~\citep{ch93,bb05}, while the whole physical picture is quite
the opposite for the fluid dynamical problem of the hydraulic
jump~\citep{bdp93,rb07,sbr05}. Contrary to all this, the axisymmetric
stationary quasi-viscous flow is greatly disturbed both by a standing
wave and by a travelling wave~\citep{br07}. 
This provides convincing evidence of
its unstable character, and it is very much in consonance with similar
conclusions drawn from some earlier studies. For high-frequency radial
perturbations~\citet{ct93} have found that inertial-acoustic modes are
locally unstable, with a greater degree of growth for the outward
travelling modes than the inward ones. On the other hand,~\citet{kat88}
have revealed a growth in the amplitude of a non-propagating perturbation
at the critical point, which, however, becomes stable in the inviscid 
regime.

This kind of instability --- one that manifests itself only if
some dissipative mechanism (viscous dissipation in the case of the
quasi-viscous rotational flow) is operative --- is called
{\em secular instability}~\citep{sc87}. It should be very much instructive
here to furnish a parallel instance of the destabilising influence of
viscous dissipation in a system undergoing rotation: that of the effect
of viscous dissipation in a Maclaurin spheroid~\citep{sc87}.
In studying ellipsoidal figures of equilibrium,~\citet{sc87}
has discussed that a secular instability develops in a
Maclaurin spheroid, when the stresses derive from an ordinary viscosity
which is defined in terms of a coefficient of kinematic viscosity (as
the $\alpha$ parametrisation is for an accretion disc), and when the
effects arising from viscous dissipation are considered as small
perturbations on the inviscid flow, to be taken into account in the
first order only. It is exactly in this spirit that the ``quasi-viscous"
approximation has been prescribed for the thin accretion disc, although,
unlike a Maclaurin spheroid, an astrophysical accretion disc is an open
system.

Curiously enough, the geometry of the fluid flow also seems to be
having a bearing on the issue of stability. The same kind of
study, as has been done here with viscosity in a rotational flow, had
also been done earlier for a viscous spherically symmetric accreting
system. In that treatment~\citep{ray03b} viscosity was found to have
a stabilising influence on the system, causing a viscosity-dependent
decay in the amplitude of a linearised standing-wave perturbation.
This is quite in keeping with the understanding that the respective
roles of viscosity are at variance with each other in the two
distinctly separate cases of spherically symmetric flows and 
disc flows. While viscosity contributes to the resistance against
infall in the former case, in the latter it aids the infall process.

An important aspect of the time-dependent perturbative analysis 
presented here is that the flow has been modelled to be driven by
a general gravitational potential, $\phi$ (as opposed to the choice 
of any particular kind of mathematical form for $\phi$). Therefore, 
none of conclusions regarding secular instability will be qualified 
upon using any of the
pseudo-Newtonian potentials~\citep{pw80,nw91,abn96}, which are regularly
invoked in accretion-related literature to describe rotational flows
on to a Schwarzschild black hole. This shall be especially true
for the flow on large scales, where all pseudo-Newtonian potentials
converge to the Newtonian limit. The following treatment will bear
this out. 

To proceed with the perturbative analysis, a time-dependent 
perturbation is introduced about the stationary solutions 
of the flow variables, $v$ and $\rho$, according to the scheme,
$v(r,t) = v_0(r) + v^{\prime}(r,t)$, $\rho (r,t) = \rho_0(r) + 
\rho^{\prime}(r,t)$ and $f(r,t) = f_0(r) + f^{\prime}(r,t)$, in 
all of which, the subscript ``$0$" implies stationary values. In 
particular, the stationary solution $f_0$ is a constant, as both 
equations~(\ref{conrho}) and (\ref{coneff}) indicate. 
This constant, as it is immediately evident from a look at 
equation~(\ref{conrho}), is very much connected to the matter flow
rate, and, therefore, the perturbation $f^{\prime}$ is to be seen as 
a disturbance on the steady, constant background accretion rate. 
For spherically symmetric flows, this Eulerian perturbation scheme has 
been applied by~\citet{pso80} and~\citet{td92}, while for inviscid 
axisymmetric flows, the same method has been used equally effectively 
by~\citet{ray03a} and~\citet{crd06}.   

The definition of $f$ will lead to a linearised dependence among 
$f^{\prime}$, $v^{\prime}$ and $\rho^{\prime}$ as 
\begin{equation}
\label{effprime}
\frac{f^{\prime}}{f_0} = \left( \frac{\gamma +1}{2} \right)
\frac{\rho^{\prime}}
{\rho_0} + \frac{v^{\prime}}{v_0} ,
\end{equation}
while from equation~(\ref{conrho}), an exclusive dependence 
of $\rho^{\prime}$
on $f^{\prime}$ will be obtained as 
\begin{equation}
\label{flucden}
\frac{\prt \rho^{\prime}}{\prt t} + \beta^2
\frac{v_0 \rho_0}{f_0} \left(\frac{\prt f^{\prime}}{\prt r}\right)=0 , 
\end{equation}
with $\beta^2 = 2(\gamma +1)^{-1}$. 
Combining equations~(\ref{effprime}) and~(\ref{flucden}) will 
render the velocity fluctuations as
\begin{equation}
\label{flucvel}
\frac{\prt v^{\prime}}{\prt t}= \frac{v_0}{f_0}
\left(\frac{\prt f^{\prime}}{\prt t}+{v_0}
\frac{\prt f^{\prime}}{\prt r}\right) , 
\end{equation}
which, upon a further partial differentiation with respect to time,
will give
\begin{equation}
\label{flucvelder2}
\frac{{\prt}^2 v^{\prime}}{\prt t^2}= \frac{v_0}{f_0}
\left[\frac{\prt^2 f^{\prime}}{\prt t^2} + v_0 \frac{\prt}{\prt r}
\left(\frac{\prt f^{\prime}}{\prt t}\right) \right] . 
\end{equation}

From equation~(\ref{radmom}) the linearised fluctuating part could be
extracted as
\begin{equation}
\label{flucradmom}
\frac{\prt v^{\prime}}{\prt t}+ \frac{\prt}{\prt r}
\left( v_0 v^{\prime} + c_{\mrm{s0}}^2
\frac{\rho^{\prime}}{\rho_0}\right)+4\alpha \lambda^2 \frac{\sigma}{r^3}
\left[2 \frac{f^{\prime}}{f_0} - 2 \frac{\rho^{\prime}}{\rho_0} 
-3 \frac{v^{\prime}}{v_0} + \frac{1}{\sigma} \int \sigma 
\frac{\prt}{\prt r} \left(\frac{f^{\prime}}{f_0}\right)\, {\mrm d}r
\right] =0 , 
\end{equation}
in which $\sigma = c_{\mrm{s0}}^2/(v_0 v_{\mrm K})$ and $c_{\mrm{s0}}$ 
is the local speed of sound in the steady state. Differentiating 
equation~(\ref{flucradmom}) partially with respect to $t$,
and making use of equations~(\ref{flucden}), (\ref{flucvel}) 
and~(\ref{flucvelder2}) to substitute for all the first and second-order
derivatives of $v^{\prime}$ and $\rho^{\prime}$, will deliver the result
\begin{equation}
\label{tpert}
\frac{{\prt}^2 f^{\prime}}{\prt t^2} +2 \frac{\prt}{\prt r}
\left(v_0 \frac{\prt f^{\prime}}{\prt t} \right) + \frac{1}{v_0}
\frac{\prt}{\prt r}\left[ v_0 \left(v_0^2-
\beta^2 c_{\mrm{s0}}^2 \right) \frac{\prt f^{\prime}}{\prt r}\right] 
- 4 \alpha \lambda^2 \frac{\sigma}{v_0 r^3}\bigg[\frac{\prt f^{\prime}}
{\prt t} + \left(\frac{3\gamma -1}{\gamma + 1}\right) v_0 
\frac{\prt f^{\prime}}{\prt r} 
- \frac{1}{\sigma} \int \sigma 
\frac{\prt}{\prt r} \left(\frac{\prt f^{\prime}}{\prt t} \right)\, 
{\mrm d}r \bigg] = 0 , 
\end{equation}
entirely in terms of $f^{\prime}$. This is the equation of motion for
a perturbation imposed on the constant mass flux rate, $f_0$, and it 
shall be important to note here that the choice of a driving potential,
Newtonian or pseudo-Newtonian, has no explicit bearing on the form 
of the equation. 

With a linearised equation of motion for the perturbation having been
derived, a solution of the form 
$f^{\prime}(r,t) = g_{\omega}(r) \exp(-{\mrm i}\omega t)$ is applied 
to it. From equation~(\ref{tpert}), this will give
\begin{equation}
\label{disp1}
\omega^2 g_\omega + 2 {\mrm i} \omega \frac{\mrm d}{{\mrm d}r}
\left(v_0 g_\omega \right)
- \frac{1}{v_0} \frac{\mrm d}{{\mrm d}r}
\left[v_0 \left(v_0^2 - \beta^2 c_{\mrm{s0}}^2 \right)
\frac{{\mrm d}g_\omega}{{\mrm d}r} \right] + 4 \alpha \lambda^2 
\frac{\sigma}{v_0 r^3} \bigg[ - {\mrm i}\omega g_\omega + 
\left(\frac{3 \gamma -1}{\gamma + 1} \right) v_0 
\frac{{\mrm d}g_\omega}{{\mrm d}r} 
+ \frac{{\mrm i} \omega}{\sigma}
\int \sigma \left(\frac{{\mrm d}g_\omega}{{\mrm d}r}\right) \, 
{\mrm d}r \bigg] = 0 .
\end{equation}

The perturbation is now made to behave in the manner of a radially
travelling high-frequency wave, whose wavelength is suitably 
constrained to be small, 
i.e. it is to be smaller than any characteristic length scale in the 
system. Effectively this invokes the {\it WKB} approximation, and a 
perturbative treatment of this nature has been carried out 
before on spherically symmetric flows~\citep{pso80} and on axisymmetric 
flows~\citep{ray03a,crd06}. In both these cases the radius
of the accretor was chosen as the characteristic length scale in
question, and the wavelength of the perturbation was required to be
much smaller than this length scale. In this study of an axisymmetric 
flow driven by the gravity of a black hole, the radius of the 
event horizon could be a choice for such a length scale. As a result, 
the frequency, $\omega$, of the waves should be large. 

An algebraic rearrangement of terms in equation~(\ref{disp1}) will 
lead to an integro-differential equation of the form 
\begin{equation}
\label{gee}
{\mathcal P}\frac{{\mrm d^2} g_\omega}{{\mrm d}r^2} 
+ {\mathcal Q}\frac{{\mrm d} g_\omega}{{\mrm d}r}
- {\mathcal R}g_\omega 
+ {\mathcal T} \int g_\omega
\left(\frac{{\mrm d} \sigma}{{\mrm d}r}\right)\,{\mrm d}r =0 , 
\end{equation}
with its coefficients being given by 
\begin{displaymath}
\label{pee}
{\mathcal P} = v_0^2 - \beta^2 c_{\mrm{s0}}^2 , 
\end{displaymath}
\begin{displaymath}
\label{qu} 
{\mathcal Q} = 3 v_0 \frac{{\mrm d}v_0}{{\mrm d}r}-\frac{1}{v_0}
{\frac{\mrm d}{{\mrm d}r}}\left( v_0 \beta^2 c_{\mrm{s0}}^2\right)
- 2{\mrm{i}}\omega v_0- 2\alpha \lambda^2\left(3\gamma-1 \right) 
\frac{\beta^2 \sigma}{r^3} ,  
\end{displaymath}
\begin{displaymath}
\label{aar} 
{\mathcal R} = 2{\mrm i} \omega \frac{{\mrm d}v_0}{{\mrm d}r}
+ \omega^2  
\end{displaymath}
and
\begin{displaymath}
\label{tee}
{\mathcal T} = \frac{4 {\mrm{i}} \omega \alpha \lambda^2}{v_0 r^3} . 
\end{displaymath}

At this stage, bearing in mind the constraint that $\omega$ is large,
the spatial part of the perturbation, $g_\omega (r)$, is prescribed as
$g_\omega (r) = \exp(s)$, with the function $s$ itself being represented 
as a power series of the form
\begin{equation}
\label{pow}
s(r)=\sum_{n=-1}^{\infty}\frac{k_n(r)}{\omega^n} . 
\end{equation}
The integral term in equation~(\ref{gee}), can,
through some suitable algebraic substitutions, be recast as 
\begin{displaymath}
\label{integ}
\int g_\omega
\left(\frac{{\mrm d} \sigma}{{\mrm d}r}\right)\, {\mrm d}r
= \int \exp(s)
\left(\frac{{\mrm d} \sigma}{{\mrm d}s}\right)\, {\mrm d}s
= g_\omega (r) {\mathcal S} ,
\end{displaymath}
with $\mathcal{S}$ being given by another power series as 
\begin{equation}
\label{powess}
{\mathcal S}= \sum_{m= 1}^{\infty} \left( -1 \right)^{m+ 1} 
\frac{{\mrm d}^m \sigma}{{\mrm d}s^m} . 
\end{equation}
Following this, all the terms in equation~(\ref{gee}) can be expanded with 
the help of the power series for $g_\omega (r)$. Under the assumption 
(whose self-consistency will be justified soon) that to a leading order 
\begin{displaymath}
\label{ess}
{\mathcal S} \sim \frac{{\mrm d}\sigma}{{\mrm d}s}
\simeq \frac{{\mrm d}\sigma}{{\mrm d}r}
\left(\omega \frac{{\mrm d} k_{-1}}{{\mrm d}r}\right)^{-1} , 
\end{displaymath}
the three successive highest-order terms (in a decreasing order) involving 
$\omega$ will be obtained as $\omega^2$, $\omega$ and $\omega^0$. The 
coefficients of each of these terms are to be collected first and then 
individually summed up. This is to be followed by setting each of these
sums separately to zero, which will yield for $\omega^2$, $\omega$
and $\omega^0$, respectively, the conditions
\begin{equation}
\label{omegsq}
\left(v_0^2 - \beta^2 c_{\mrm{s0}}^2\right)
\left( \frac{{\mrm d}k_{-1}}{{\mrm d}r} \right)^2
-2 {\mrm i} v_0 \frac{{\mrm d}k_{-1}}{{\mrm d}r} -1 = 0 , 
\end{equation}
\begin{equation}
\label{omeg1}
\left(v_0^2 - \beta^2 c_{\mrm{s0}}^2\right)
\left( \frac{{\mrm d}^2 k_{-1}}{{\mrm d}r^2}
+ 2 \frac{{\mrm d}k_{-1}}{{\mrm d}r}
\frac{{\mrm d}k_0}{{\mrm d}r} \right)
+ \left[ 3 v_0 \frac{{\mrm d}v_0}{{\mrm d}r}
- \frac{1}{v_0}\frac{\mrm d}{{\mrm d}r}
\left( v_0 \beta^2 c_{\mrm{s0}}^2 \right) -2 \alpha \lambda^2 
\left(3 \gamma -1 \right) \frac{\beta^2 \sigma}{r^3} \right]
\frac{{\mrm d}k_{-1}}{{\mrm d}r}
- 2{\mrm i} v_0 \frac{{\mrm d}k_0}{{\mrm d}r}
- 2{\mrm i} \frac{{\mrm d}v_0}{{\mrm d}r} = 0
\end{equation}
and
\begin{displaymath}
\label{omeg0}
\left( v_0^2 - \beta^2 c_{\mrm{s0}}^2 \right)
\left[ \frac{{\mrm d}^2 k_0}{{\mrm d}r^2}
+ 2 \frac{{\mrm d}k_{-1}}{{\mrm d}r}
\frac{{\mrm d}k_1}{{\mrm d}r} +
\left( \frac{{\mrm d}k_0}{{\mrm d}r} \right)^2
\right] + 
\left [3 v_0 \frac{{\mrm d}v_0}{{\mrm d}r}
- \frac{1}{v_0} \frac{\mrm d}{{\mrm d}r}
\left(v_0 \beta^2 c_{\mrm{s0}}^2 \right) 
- 2 \alpha \lambda^2 \left(3 \gamma -1\right)\frac{\beta^2 \sigma}
{r^3} \right]\frac{{\mrm d}k_0}{{\mrm d}r}
\end{displaymath}
\begin{equation}
\label{contdomeg0}
\qquad \qquad \qquad \qquad \qquad \qquad \qquad \qquad
\qquad \qquad \qquad \qquad \qquad \qquad \qquad \qquad
- 2{\mrm{i}}{v_0}\frac{{\mrm d}k_1}{{\mrm d}r}
+ 4 {\mrm i} \alpha \lambda^2 \frac{1}{v_0 r^3} 
\frac{{\mrm d}\sigma}{{\mrm d}r}
\left(\frac{{\mrm d} k_{-1}}{{\mrm d}r}\right)^{-1} = 0 . 
\end{equation}
Out of these, the first two, i.e. equations~(\ref{omegsq}) 
and~(\ref{omeg1}), will deliver the solutions
\begin{equation}
\label{kayminus1}
k_{-1} = \int \frac{{\mrm{i}}}{v_0 \pm \beta c_{\mrm{s0}}} \,
{\mrm d}r
\end{equation}
and
\begin{equation}
\label{kaynot}
k_0 = - \frac{1}{2} \ln \left( v_0 \beta c_{\mrm{s0}} \right)
\pm \alpha \lambda^2 \left(3 \gamma -1 \right) \int
\frac{\beta c_{\mrm{s0}}\left(v_0 \pm \beta c_{\mrm{s0}}\right)}
{v_0 v_{\mrm{K}} r^3\left(v_0^2 - \beta^2 c_{\mrm{s0}}^2\right)} \,
{\mrm{d}}r , 
\end{equation}
respectively. 

The two foregoing expressions give the leading terms in the power 
series of $g_\omega (r)$. While dwelling on this matter, it will
also be necessary to show that all successive terms of $s(r)$ will
self-consistently follow the condition
$\omega^{-n}\vert k_n(r)\vert \gg \omega^{-(n+1)}\vert k_{n+1}(r)\vert$,
i.e. the power series given by $g_\omega (r)$ will converge very quickly 
with increasing $n$. In the inviscid limit, this requirement can be shown 
to be very much true, considering the behaviour of the first three 
terms in $k_n(r)$ from equations~(\ref{kayminus1}), (\ref{kaynot})
and~(\ref{contdomeg0}). These terms can be shown to go
asymptotically as
$k_{-1} \sim r$, $k_0 \sim \ln r$ and $k_1 \sim r^{-1}$, given the
condition that $v_0 \sim r^{-5/2}$ on large length scales, while
$c_{\mrm{s0}}$ approaches its constant ambient value. With the 
inclusion of viscosity as a physical effect, it can be seen from
equations~(\ref{kayminus1}) and~(\ref{kaynot}), respectively, that while
$k_{-1}$ remains unaffected, $k_0$ acquires an $\alpha$-dependent
term that goes asymptotically as $r$. This in itself is an
indication of the extent to which viscosity might alter the inviscid
conditions. However, since $\alpha$ has been chosen to be very much
less than unity, and since the wavelength of the travelling waves 
is also very small, 
the self-consistency requirement still holds. Therefore, as far
as gaining a qualitative understanding of the effect of viscosity 
is concerned, it should be quite sufficient to truncate the power
series expansion of $s(r)$, after considering the two leading terms 
only, and with the help of these two, an expression for the 
perturbation may then be set down as
\begin{equation}
\label{fpertur}
f^{\prime}(r,t) \simeq \frac{A_\pm}{\sqrt{\beta v_0 c_{\mrm{s0}}}}
\exp \left[ \pm \alpha \lambda^2 \left(3 \gamma -1 \right) \int 
\frac{\beta c_{\mrm{s0}}\left(v_0 \pm \beta c_{\mrm{s0}}\right)}
{v_0 v_{\mrm{K}} r^3\left(v_0^2 - \beta^2 c_{\mrm{s0}}^2\right)} 
\, {\mrm d}r \right] 
\exp \left ( \int \frac{{\mrm i} \omega}{v_0 \pm \beta
c_{\mrm{s0}}} \, {\mrm d}r \right)
e^{-{\mrm i} \omega t} , 
\end{equation}
which should be seen as a linear superposition of two waves with 
arbitrary constants $A_+$ and $A_-$. Both of these two waves move 
with a velocity $\beta c_{\mrm{s0}}$ relative to the fluid,
one against the bulk flow and the other along with it, while the bulk
flow itself has a velocity $v_0$. It should be immediately evident 
that all questions pertaining to the growth or decay in the amplitude of
the perturbation will be crucially decided by the real terms delivered
from $k_0$. The viscosity-dependent term is especially crucial
in this regard. For the choice of the lower sign in the real part of
$f^\prime$ in equation~(\ref{fpertur}), i.e. for the outgoing mode of the
travelling wave solution, it can be seen that the presence of viscosity 
causes the amplitude of the perturbation to diverge exponentially on 
large length scales, where $c_{\mrm{s0}} \simeq c_{\mrm{s}}(\infty)$ 
and $v_0 \sim r^{-5/2}$, with $-v_0$ being positive for inflows.
The inwardly travelling mode also displays similar behaviour, albeit
to a quantitatively lesser degree.
It is an easy exercise to see that stability in the system would be
restored for the limit of $\alpha = 0$, and this particular issue has
been discussed by~\citet{ray03a} and~\citet{crd06}. The exponential growth 
behaviour of the amplitude of the perturbation, therefore, is exclusively 
linked to the presence of viscosity. Going back to a work 
of~\citet{ct93}, it can be seen that the inertial-acoustic modes of short 
wavelength radial perturbations are locally unstable throughout
the disc, with the outward travelling modes growing faster than the
inward travelling modes in most regions of the disc, all of which
is very much in keeping with what equation~(\ref{fpertur}) indicates here.

With the help of equation~(\ref{flucden}) it should be easy to express 
the density fluctuations in terms of $f^{\prime}$ as
\begin{equation}
\label{effden}
\frac{\rho^{\prime}}{\rho_0} = \beta^2 \left(\frac{v_0}{{\mrm i}\omega}
\frac{{\mrm d} s}{{\mrm d} r} \right)\frac{f^{\prime}}{f_0} , 
\end{equation}
and likewise, the velocity fluctuations may be set down from 
equation~(\ref{effprime}) as 
\begin{equation}
\label{effvel}
\frac{v^{\prime}}{v_0} = \left(1 - \frac{v_0}{{\mrm i}\omega}
\frac{{\mrm d} s}{{\mrm d} r} \right)\frac{f^{\prime}}{f_0} . 
\end{equation}

In a unit volume of the fluid, the kinetic energy content is
\begin{equation}
\label{ekin}
{\mathcal E}_{\mrm{kin}} = \frac{1}{2} \left(\rho_0
+ \rho^{\prime}\right)\left(v_0 + v^{\prime}\right)^2 , 
\end{equation}
while the potential energy per unit volume of the fluid is the sum of
the gravitational energy, the rotational energy and the internal energy.
For a quasi-viscous disc, to a first order in $\alpha$, this sum is 
given by
\begin{equation}
\label{epot}
{\mathcal E}_{\mrm{pot}} = \left(\rho_0 + \rho^{\prime}\right) 
\left[ \phi (r) + \frac{\lambda_{\mrm{eff}}^2}{2r^2} \right]
+ \rho_0 \epsilon + \rho^{\prime} \left[\frac{\prt}{\prt \rho_0}
\left(\rho_0 \epsilon \right)\right] + \frac{1}{2}{\rho^{\prime}}^2
\left[ \frac{{\prt}^2}{\prt \rho_0^2}\left(\rho_0 \epsilon \right) \right] , 
\end{equation}
where $\epsilon$ is the internal energy per unit mass~\citep{ll87}.
In equation~(\ref{epot}) the effective angular momentum for the 
quasi-viscous disc will have to be set up as a first-order correction
about the inviscid conditions. Following this, a time-dependent 
perturbation has to be imposed about the stationary values of $v$ 
and $\rho$. All first-order terms involving time-dependence in  
equations~(\ref{ekin}) and~(\ref{epot}) will 
vanish on time-averaging. In this situation the leading contribution 
to the total energy in the perturbation comes from the second-order 
terms, which are all summed as 
\begin{displaymath}
\label{order2}
{\mathcal E}_{\mrm{pert}} = \frac{1}{2} \rho_0 {v^{\prime}}^2
+ v_0 \rho^{\prime} v^{\prime} +  \frac{1}{2} {\rho^{\prime}}^2
\left[\frac{{\prt}^2}{\prt \rho_0^2} \left(\rho_0 \epsilon \right) \right]
- 2 \alpha \lambda^2 \frac{\rho_0 \sigma}{r^2} \bigg[ \left(
\frac{\rho^{\prime}}{\rho_0}\right)^2 + \left(\frac{f^{\prime}}
{f_0}\right)^2 + 6 \left(\frac{v^{\prime}}{v_0}\right)^2
- 2 \frac{\rho^{\prime}f^{\prime}}{\rho_0 f_0}
+ 3 \frac{v^{\prime} \rho^{\prime}}{v_0 \rho_0} 
- 6 \frac{f^{\prime} v^{\prime}}{f_0 v_0} 
\end{displaymath}
\begin{equation}
\label{contdorder2}
\qquad \qquad \qquad \qquad \qquad \qquad \qquad \qquad \qquad
+ \frac{1}{\sigma} \left(\frac{\rho^{\prime}}{\rho_0} \right) \int 
\sigma \frac{\mrm d}{{\mrm d} r} \left(\frac{f^{\prime}}{f_0} \right)\, 
{\mrm d}r + \frac{1}{\sigma} \int \sigma \left(\frac{f^{\prime}}{f_0}
- 2 \frac{\rho^{\prime}}{\rho_0} - 3 \frac{v^{\prime}}{v_0} \right)
\frac{\mrm d}{{\mrm d} r} \left(\frac{f^{\prime}}{f_0} \right)\,
{\mrm d}r \bigg] . 
\end{equation}
In the preceding expression all terms involving $\rho^{\prime}$ and 
$v^{\prime}$ can be written in terms of $f^{\prime}$ with the help of
equations~(\ref{effden}) and~(\ref{effvel}), in both of which, to a 
leading order, $s \simeq \omega k_{-1}$. This is to be followed 
by a time-averaging over ${f^{\prime}}^2$, which will  
contribute a factor of $1/2$. The total energy flux in the perturbation 
is obtained by multiplying ${\mathcal E}_{\mrm{pert}}$
by the propagation velocity $(v_0 \pm \beta c_{\mrm{s0}})$ and then
by integrating over the area of the cylindrical face of the accretion
disc, which is $2 \pi rH$. Under the thin-disc approximation, $H \ll r$,
this will make it possible to derive an estimate for the energy flux as 
\begin{equation}
\label{flux}
{\mathcal F} (r) \simeq 
\frac{\pi \beta^2 A_{\pm}^2 \sqrt{K}}{f_0}
\left[\pm 1 + \frac{1- \beta^2\left(2 - \mu \right)}
{2 \beta \left({\mrm M} \pm \beta \right)}\right]
\left(1 - \frac{2 \alpha \lambda^2 \zeta}
{\beta^2 {\mrm M}^2 v_0 v_{\mrm K}r^2} \right) 
\exp \left[ \pm 2 \alpha \lambda^2 \left(3 \gamma -1 \right) \int
\frac{\beta c_{\mrm{s0}}\left(v_0 \pm \beta c_{\mrm{s0}}\right)}
{v_0 v_{\mrm{K}} r^3\left(v_0^2 - \beta^2 c_{\mrm{s0}}^2\right)}
\, {\mrm d}r \right] , 
\end{equation}
in which ${\mrm M}$ is the Mach number, defined as 
${\mrm M} = v_0/c_{\mrm{s0}}$, while 
\begin{displaymath}
\label{mu}
\mu = \frac{\rho_0}{c_{\mrm{s0}}^2} \left[
\frac{\prt^2\left(\rho_0 \epsilon\right)}{\prt \rho_0^2} \right] 
\end{displaymath}
and 
\begin{displaymath}
\label{exppsi}
\zeta = 2 \beta \left[\left(\beta^2 -1 \right)^2 {\mrm M}^2
\pm \beta {\mrm M} \left(\beta^2 - 4 \right) + \beta^2 \right]
\left[1 \pm 2 \beta {\mrm M} + \beta^2 \mu\right]^{-1} . 
\end{displaymath}
When ${\mrm M} \longrightarrow 0$ on large length scales, $\zeta$ 
converges to a finite value. However, on these same length 
scales, what will {\em not} converge are the two terms involving $\alpha$
in equation~(\ref{flux}). Under the asymptotic conditions on $v_0$ and 
$c_{\mrm{s0}}$, discussed earlier, one term will diverge exponentially,  
while another will have a power-law growth behaviour of $r^6$. 
The quasi-viscous disc will, therefore, be unstable 
on large scales under the passage of a linearised radially propagating 
high-frequency perturbation. 
It is easy to check that under inviscid conditions, with $\alpha =0$,  
and for an adiabatic perturbation with $\mu =1$, the disc will 
immediately revert to stable behaviour, a feature that has been 
mathematically demonstrated in some earlier 
papers~\citep{ray03a,crd06}.
That instabilities should develop exponentially on large length 
scales, because of the merest presence of viscosity (to a first order 
in $\alpha$, which itself is much less than unity) has disturbing 
implications, since all physically meaningful inflow solutions
will have to pass through these length scales, connecting the outer
boundary of the flow with the surface of the accretor (or the event
horizon, if the accretor is a black hole).

\section{Asymptotic behaviour of the quasi-viscous disc}
\label{sec5}

A previous study~\citep{rbcqg07} has dwelt on how it should be 
possible to select the stationary transonic solution 
of the inviscid axisymmetric flow through a non-perturbative
time-dependent criterion (such as the minimisation of the total 
specific energy of the flow). This kind of insight into 
the long-time behaviour of the inviscid flow on large length scales 
of the thin accretion disc allows for making 
a similar foray into the quasi-viscous disc. However, in this
instance, a straightforward application of the methods employed
for the inviscid disc would not be possible on two counts. First, the
quasi-viscous disc being a dissipative system (i.e. with energy being 
allowed
to be drained away from this system), there should be no occasion to
look for the selection of a particular solution, and a selection
criterion thereof, on the basis of energy minimisation. Secondly,
the fact that the quasi-viscous disc is unstable on large length 
scales, is reason enough to believe that no solution
--- transonic or otherwise --- might be free of time-dependence. 
Therefore, a long-time evolution of the quasi-viscous disc towards 
a stationary end is not something that might be hoped for. For all
that, however, this kind of a disc system does exhibit some interesting
asymptotic features on large length scales. 

On such length scales of an accretion disc, all pseudo-Schwarzschild
flows converge to the Newtonian limit, i.e. $\phi (r) \sim - GM/r$. 
Meanwhile the stationary solution of equation~(\ref{conrho}) can 
be expressed on the same length scales simply as,
\begin{equation}
\label{statiocon}
\rho_0^{(\gamma +1)/2} v_0 r^{5/2} \sim -\dot{m} \sqrt{GM},
\end{equation}
with $\dot{m}$ being the conserved matter inflow rate. The negative sign
arises because for inflows, $v_0$ goes with a negative sign.
From equation~(\ref{statiocon}), with $\rho_0$ approaching a constant
ambient value on large length scales, the drift velocity, $v_0$, can
consequently be seen to go asymptotically as $r^{-5/2}$. Bearing in 
mind that for inflows, $v_0 <0$, the asymptotic dependence of the 
effective angular momentum can be shown from equation~(\ref{statioang})
to be
\begin{equation}
\label{asympang}
\lambda_{\mrm{eff}}(r) \sim \lambda \left[ 1 + 2 \alpha 
\left(\frac{r}{r_{\mrm l}}\right)^3 \right] ,
\end{equation}
where $r_{\mrm l}$ is a scale of length, which, to an
order-of-magnitude, is given by
$r_{\mrm l}^3\sim GM{\dot{m}}[c_{\mrm s}^3(\infty)\rho(\infty)]^{-1}$.
This asymptotic behaviour is entirely to be expected, because the
physical role of viscosity is to transport angular momentum to large
length scales of the accretion disc.

Nevertheless, the distribution of matter in a viscous disc takes place
on a time scale determined by viscosity, and therefore a study of the
time-dependent properties in a viscous disc is one of the few means
of acquiring some impression about the role of viscosity, especially
since the observables in a steady disc are largely independent of
viscosity~\citep{fkr02}. With that objective in mind, it will be 
worthwhile first to try to understand the structure of the governing
time-dependent differential equation for the flow on large length
scales. To do so, it shall be necessary to invoke the approximation 
that very far from the accretor in the outer regions of the flow, 
on highly subsonic scales of velocity, the
density variations are negligibly small compared to the time evolution
of the velocity field. The evolution will consequently follow the 
general Navier-Stokes equation in the limit of $\prt_j v_j = 0$, 
which can be set down as 
\begin{equation}
\label{velfield}
\prt_t v_i + v_j \prt_j v_i = \nu \prt_j \prt_j v_i
- \prt_i V , 
\end{equation}
where the potential function $V \equiv V(r,t) = nc_{\mrm s}^2 
+ \phi (r) + \lambda^2/2 r^2$, with $n$ being the usual polytropic
index. Quite evidently, equation~(\ref{velfield}) is a nonlinear 
differential equation, but with the help of the Hopf-Cole 
transformation~\citep{regev}, 
\begin{equation}
\label{hopf-cole}
v_i = - \frac{2 \nu}{\xi} \prt_i \xi ,
\end{equation} 
it can be reduced to a linear form in the variable $\xi$, 
going as 
\begin{equation}
\label{linphi}
2 \nu \prt_t \xi = 2 \nu^2 \prt_i \prt_i \xi + V \xi . 
\end{equation}
The potential function $V$ will in general be modified by the addition 
of an integration constant, which might physically be identified
from the ambient conditions of the fluid. This, however, will 
require a knowledge of the boundary conditions for the scalar function
$\xi$, something whose inherent difficulties would be appreciated soon.
In the outer regions of the disc, however, the speed of sound, which is
a scalar function of the density, would
asymptotically assume a constant value.

An interesting aspect of equation~(\ref{linphi}) is that for the
one-to-one correspondence of $2 \nu$ with ${\mrm i} \hbar$, and 
of $m$ with $1$, there is an exact equivalence between this equation 
and Schr\"odinger's equation, 
\begin{equation}
\label{schro}
{\mrm i}\hbar \prt_t \psi =-
\frac{\hbar^2}{2m}\prt_i \prt_i \psi +V\psi . 
\end{equation} 
For the steady limit of the potential function $V$ (a requirement that
is satisfied on large length scales), it is easy to carry out a separation 
of variables in equation~(\ref{linphi}), and the resulting stationary
eigenvalue equation in $\xi$ may then be expressed in a Hamiltonian 
form as
\begin{equation}
\label{hamil}
- 2 \nu^2 \nabla^2 \xi + \tilde{V} \xi
= E \xi , 
\end{equation}
in which $\tilde{V}=-V$. A comparison between  
equation~(\ref{hamil}) and the stationary Schr\"odinger equation
ought to be instructive and insightful here. 
To have any notion of how $\xi$
evolves in time, one would have to determine the eigenvalues given
by $E$, and it is here that a great stumbling block is
encountered.
To solve the second-order differential equation given 
by equation~(\ref{hamil}),
two boundary conditions are imperative, and they in turn would
characterise the eigenvalues. The outer boundary condition on $\xi$
is relatively easy to prescribe. For $r \longrightarrow \infty$ and
$v\longrightarrow 0$, the scalar function $\xi$ will asymptotically
approach a constant value, and there is much similarity in this with
the asymptotic behaviour of another scalar function, the density
$\rho$. Knowing the precise inner boundary condition on $\xi$, however,
is a most difficult problem. First of all, it depends on the nature of
the accretor itself. While a black hole will have an event horizon,
a neutron star or an ordinary star will have a physical surface. This
fact alone has much influence over the inner boundary condition. Apart 
from this, realistically speaking, various astrophysical processes near 
the surface of the accretor will affect the flow~\citep{pso80}. In any 
case, equation~(\ref{hamil}) is valid for large scales only. Yet, short 
of actually having to solve equation~(\ref{hamil}), it should still be 
possible to derive some information
on how viscosity affects the flow in the outer regions of the disc.
It is important to see that $\tilde{V}$ in equation~(\ref{hamil})
assumes the properties of a repulsive
potential. More than the pressure effects, this
repulsive nature is reflective of the cumulative transfer of angular
momentum to large length scales of an accretion disc. Referring to
equation~(\ref{asympang}), one can see that in the outer regions of
the disc, the effective specific force, $\varphi$, is given to a 
first order in $\alpha$ by
\begin{equation}
\label{effforce}
\varphi (r) \sim - \frac{GM}{r^2} + \frac{\lambda^2}{r^3}
+ 4\alpha \frac{\lambda^2}{r_{\mrm l}^3} , 
\end{equation}
from which it is evident that on scales of $r \sim r_{\mrm l}$, the 
transport
of angular momentum will give rise to an asymptotic constant non-zero
force opposed to gravity. For the viscous disc this gives rise to a 
repulsive effect on large length scales, in opposition to gravitational
attraction. From this argument one may go a step beyond and conjecture 
that $r_{\mrm l} \alpha^{-1/3}$ defines a limiting length scale for 
accretion, that the outward transport of angular momentum imposes. 
This is also the length scale on which secular instability is most 
conspicuous. In units of the Schwarzschild radius of the black 
hole, $2GM/c^2$, this length scale is seen to be
\begin{equation}
\label{schawscale}
r_{\mrm{Sl}} \sim \frac{c^2}{2 c_{\mrm s}(\infty)}
\left[\frac{\dot{m}}{\alpha (GM)^2 \rho(\infty)} \right]^{1/3} .
\end{equation}
All of this is quite compatible with how viscosity redistributes
an annulus of matter in a Keplerian flow around an accretor;
the inner region drifting in because of dissipation, and consequently,
through the conservation of angular momentum and its outward transport,
making it necessary for the outer regions of the matter distribution
to spread even further outwards~\citep{pri81,fkr02}. This state of
affairs is qualitatively not altered in anyway for the quasi-viscous
flow, except
for the fact that with viscosity being very weak here, the outward
transport of angular momentum can perceptibly cause an outward drift
of matter only on very large scales. It is obvious that once
$\alpha =0$, i.e. for the inviscid limit, as equation~(\ref{schawscale})
shows, this scale would be shifted to infinity.

\section{Concluding remarks}
\label{sec6}

One very important physical role of viscosity in an accretion disc
is that it governs the distribution of matter in the disc. 
The manner in which viscosity redistributes an annulus of matter 
in a Keplerian flow around an accretor is very well known, with 
the inner region of this disc system drifting in because of dissipation, 
and consequently making it necessary for the outer regions of the 
matter distribution to spread out even farther, because of the
conservation of angular momentum and its outward 
transport~\citep{pri81,fkr02}. 

Viscosity, however, also gives rise to secular instability 
in the quasi-viscous disc. This casts much 
doubt on the long-term viability of the accretion flow, 
and its temporal evolution towards a stationary state. 
It may rightly be argued that the instability that develops on 
the large subsonic scales of a quasi-viscous disc is intimately
connected with the cumulative transfer of angular momentum on 
these very length scales. The accumulation of angular momentum in 
this region may create an abrupt centrifugal barrier against any 
further smooth inflow of matter. However, this adverse effect 
could disappear if there
should be some other means of transporting angular momentum from 
the inner regions of the disc. Astrophysical jets could readily
afford such a means~\citep{witjet}, insofar as jets actually cause 
a physical 
drift of angular momentum vertically away from the plane of the 
disc, instead of along its radial length. This will be all the 
more true if this off-the-plane angular momentum drift happens 
on length scales that are much smaller than the scale indicated 
by equation~(\ref{schawscale}). 

Stability could be restored through many other means. A recent work
by~\citet{macmal08} has shown numerically how it should be possible 
to have stable steady accretion solutions for transonic flows of a 
self-gravitating gas. This stability argument holds true even for 
perturbations in the nonlinear regime. Another work 
by~\citet{nagyam08} has established the stability of an accretion shock
(connecting transonic solutions passing through two distinct saddle
points) under axisymmetric perturbations. 

\section*{Acknowledgements}

This research has made use of NASA's Astrophysics Data System. 
The authors express their gratitude to Rajaram Nityananda and 
Paul J. Wiita for some helpful comments. AKR would like to 
acknowledge the kind hospitality provided by HRI, Allahabad, India. 
The work of TKD was partially supported by  the Theoretical
Institute for Advanced Research in Astrophysics (TIARA)
operated under Academia Sinica and the National Science
Council Excellence Projects program in Taiwan, administered 
through grant NSC 96-2752-M-007-007-PAE.

\end{document}